\documentstyle[12pt,world_sci,epsf]{article}
\begin{document}
\hfill{ANL-HEP-CP-94-53}
\vskip .5cm
\hfill{ITP-SB-94-40}
\vskip 1cm

\title{{\bf PRINCIPAL-VALUE RESUMMATION \\
FOR DILEPTON PRODUCTION}}
\author{HARRY CONTOPANAGOS
%\footnote{Supported by the U.S. Department of Energy,
% Contract W-31-109-ENG-38.}
\\
{\em High Energy Physics Division, Argonne National Laboratory, \\
Argonne, IL 60439-4815, USA}\\
\vspace{0.3cm}
and\\
\vspace*{0.3cm}
LYNDON ALVERO, GEORGE STERMAN\\
{\em Institute for Theoretical Physics, SUNY Stony Brook, \\
Stony Brook, NY 11794-3840, USA}}

\maketitle
\setlength{\baselineskip}{2.6ex}

\begin{center}
\parbox{13.0cm}
{\begin {center} ABSTRACT \end{center}
{\small \hspace*{0.3cm} We present a new method of resummation of
QCD threshold distributions to the hard scattering
function of the dilepton production cross section in hadron-hadron
collisions. Our formulation
by-passes the infrared singularities of the QCD running coupling
through a principal value prescription, and does not
require an explicit infrared cut-off. The resulting
large corrections exponentiate, and
we discuss asymptotic properties of the exponent.
We present a closed analytical formula for the resummed hard
part in momentum space,
and give predictions, at fixed-target energies,
for the resulting cross section
which
is a bounded function of its kinematic variables in the
entire partonic phase space.}}
\end{center}

\section{Introduction}

It is well-known that perturbative QCD (pQCD) corrections to certain
hadron-hadron scattering cross sections can be numerically important even at
quite high energies. One source of
large corrections is soft-gluon radiation
near partonic threshold.   The best-studied example
of this mechanism is
 the dilepton-production cross section, which is not only of
great phenomenological importance but is also a true laboratory for
testing various ideas on pQCD.  As we shall see, at fixed-target
energies, the resummed series can differ significantly from
even two-loop expansions.
Indeed,
the perturbative series evaluated order-by-order
does not appear to converge, and in fact, due to the presence of
infrared (IR) renormalons, is at best only asymptotic.
This suggests that predictions of
the normalization of such cross sections on the basis of
low-order calculations combined with
parton distributions determined from deeply inelastic
scattering are unreliable.

Resummation formulas typically organize the large perturbative
corrections in an exponential containing integrals of the QCD running
coupling $\alpha_s(\mu^2)$ over various momentum scales. The
resulting expressions, however,
suffer from divergences$^1$ associated with the
singularity in the perturbative coupling at $\Lambda_{\rm QCD}$.
One way to deal with this problem is
to introduce an explicit infrared cutoff.
The dependence on such a cutoff is higher-twist$^1$.
The resulting dependence on the cutoff, however, is
difficult to analyze, and is not negligible in general.

Here we summarize recent work$^{2,3,4,5}$
 that advances a new resummation method of large
threshold corrections, which we have
called principal-value resummation.
In this approach, the integral over
the momentum scales of the running coupling
in the resummed exponent is defined as
a principal value integral, thus by-passing the IR singularities
of the coupling.
It can therefore be taken as a mathematically unambiguous
definition of the resummed perturbative series.  This then allows us to
define and explore the structure of
its higher-twist corrections, since they are
defined relative to a well-defined perturbative expression.
Furthermore, the new resummed exponent exhibits the behavior
anticipated in pQCD:
at each point in partonic phase space
it is best-approximated by
a particular
polynomial in $\alpha_s$.
It also becomes a purely non-perturbative, higher-twist
expression in the non-perturbative region
of the partonic phase space.
The resulting hard part
is calculable numerically as well as analytically (including inversion
of the Mellin transform).
We will summarize all of the above properties within
the context of the Drell-Yan process.

\section{Principal-Value Resummation}

We will focus on the reaction
\begin{equation}
h_1(p_1)+h_2(p_2)\to l{\bar l}(Q) +X,
\label{one}
\end{equation}
where in the following we use the notation
$s=(p_1+p_2)^2,$ and $z=\tau/x_ax_b$, with
$x_a,\ x_b$ the momentum fractions of the scattered partons.
We will concentrate on the diagonal-flavor quark-antiquark subprocess which
is singular in the soft-gluon limit $z\to 1$ at fixed order of pQCD.

The corresponding cross section may be written as
\begin{equation}
{d\sigma\over dQ^2}=\sigma_0^V(Q^2)W^V(\tau,Q^2),
\label{two}
\end{equation}
where $\sigma_0^V$ is the point-like tree-level cross section for
$q{\bar q}\to V^*\to l{\bar l}$, containing all electroweak parameters and
physical dimensions and
\begin{equation}
W^V(\tau,Q^2)=\sum_f C_f^V\int_0^1{dx_a\over x_a}{dx_b\over x_b}F_{f/h_1}(x_a,
Q^2)F_{{\bar f}/h_2}(x_b,Q^2)\omega_{f{\bar f}}(\tau/x_ax_b,\alpha),
\label{three}
\end{equation}
where $\alpha\equiv\alpha_s(Q^2)/\pi$ and $C_f^V$, in units of $|e|$,
is the coupling constant of parton $f$ to vector boson $V$.
Specifically, $C_f^\gamma=e_f^2$ and
$C_f^Z=1+(1-4|e_f|\sin^2\theta_W)^2$.   The sum in eq.~(\ref{three})
runs over all relevant quark and antiquark flavors and the
parton distributions $F_{f/h_1},\ F_{{\bar f}/h_2}$ must be in DIS scheme form.

For the Mellin transform of the hard part,\ $\omega_{f{\bar f}}(z,\alpha)$,
we obtain the principal-value resummation formula in moment space$^3$
\begin{equation}
\tilde{\omega}_{f{\bar f}}(n,\alpha)=A(\alpha){\rm e}^{E(n,\alpha)},
\label{four}
\end{equation}
where the function $A(\alpha)$ exponentiates large ``constants"$^6$, i.e.,
terms that depend only on $\alpha$ and not on the
parton momenta, and the resummation exponent $E(n,\alpha)$
depends on the soft gluon momenta through the correspondence
$n\leftrightarrow 1/(1-z)$. It is this quantity that we will define
in principal-value resummation.

%\subsection{The resummation exponent}

We may write$^1$
\begin{equation}
\Gamma_{f\bar{f}}(1-\zeta,Q^2)\equiv\int_0^\zeta{dy\over 1-y}g_1(\alpha[
(1-\zeta)(1-y)Q^2])+g_2(\alpha[(1-\zeta)Q^2]),
\label{five}
\end{equation}
and define the principal-value exponent as$^3$
\begin{equation}
E(n,Q^2)=-\int_Pd\zeta{\zeta^{n-1}-1\over 1-\zeta}\Gamma_{f{\bar f}}(1-\zeta,
Q^2),
\label{six}
\end{equation}
where $P\equiv {1\over 2}\times C_+\ \cup \ {1\over 2}\times C_-,
\ {\rm with}\  C_+(C_-)$
any contour in the $\zeta$ plane connecting 0 and 1 above (below) the
real axis and with $\alpha(\lambda Q^2)$
 truncated to only those terms that create
the large threshold corrections,
\begin{equation}
\alpha(\lambda Q^2)\equiv\alpha(\lambda)={\alpha\over 1+\alpha b_2\ln\lambda}
-{\alpha^2b_3\over b_2}{\ln(1+\alpha b_2\ln\lambda)\over
(1+\alpha b_2\ln\lambda)^2},
\label{seven}
\end{equation}
where $b_2={1\over 12}(33-2n_f)$ and $b_3={1\over 48}(306-38n_f)$,
$n_f$ being the number of active flavors.
The functions $g_i,\ i=1,2$ in eq.~(\ref{five}) are simple expansions
of the form $g_i(\alpha(\lambda))=\sum_{j=1}^\infty \alpha^j(\lambda)
g_i^{(j)}$ and the corresponding coefficients are factorization
scheme- and process-dependent and can be fitted to finite-order results.
What makes the procedure finite and
useful is that we {\it only need} three of these coefficients to
account for {\it all} large threshold corrections to all orders.
These, in the DIS scheme, are
\begin{equation}
g_1^{(1)}=2C_F,\ g_2^{(1)}=-{3\over 2}C_F,\ g_1^{(2)}=
C_F\biggl[C_A\biggl({67\over 18}-
{1\over 6}\pi^2\biggr)-{5\over 9}n_f\biggr].
\label{eight}
\end{equation}
It can be shown$^4$ that the {\it purely} perturbative regime is defined,
in moment space, by the relation
\begin{equation}
1<\ln n<{1\over 2\alpha b_2}\leftrightarrow 1\ll n<{Q\over \Lambda},
\label{nine}
\end{equation}
where, in the second relation,
we have used for simplicity a 1-loop relation between
$\alpha$ and $\Lambda\equiv\Lambda_{QCD}$.
In this perturbative range, a small-$\alpha$ asymptotic approximation
leads to the perturbative formula for the exponent
\begin{equation}
E(n,\alpha)\simeq E(n,\alpha,{\cal N})=\sum_{i=1}^2\sum_{\rho=1}^{N_i+1}
\alpha^\rho\sum_{j=0}^{\rho+1}s_{j,\rho}(i)\ln^jn+
\sum_{i=3}^7\sum_{\rho=1}^{N_i+1}\alpha^\rho\sum_{j=0}^\rho s_{j,\rho}(i)
\ln^jn,
\label{ten}
\end{equation}
where the $\{s_{j,\rho}(i)\}$ are calculable numerical coefficients and
${\cal N}={\cal N}(Q)=\{N_i\}$, the set of optimum numbers of perturbative
terms, is determined by
optimization of the perturbative asymptotic approximation, eqs.~(\ref{six}),
(\ref{ten}). At typical fixed-target dilepton mass scales,
say $Q=10$ GeV, we find$^{3,4}$ ${\cal N}=\{2,6,6,2,6,1,6\}$.

It can similarly be shown$^4$ that the {\it purely}
non-perturbative regime is defined
through the relation
\begin{equation}
\ln n>{1\over \alpha b_2}\leftrightarrow n>{Q^2\over \Lambda^2}.
\label{eleven}
\end{equation}
Exact analytical results for the exponent in this
region have been obtained$^4$ through a large-$\alpha$ asymptotic
approximation, but
suffice it to say that for extremely large values of $n$,
\begin{equation}
E(n,\alpha)\simeq -{g_1^{(1)}\over b_2}\ln 2\ln n-{g_2^{(1)}\over b_2}\ln(
\alpha b_2\ln n)-{g_1^{(1)}b_3\over 4b_2^3}\ln^2(\alpha b_2\ln n).
\label{twelve}
\end{equation}
This shows that the exponent tends to $-\infty$ at the edge of
phase space $z\simeq 1$ ({\it i.e.}, $n\rightarrow \infty$).
As a result, the
corresponding resummed cross section is finite, because
the  hard part is exponentially damped at that threshold.
This results in the principal-value cross section being a bounded function
of $\tau$ throughout its kinematic range, in contrast to the finite-order
cross section that diverges as a power of the logarithm as $\tau\to 1$.

To obtain the resummed hard part in momentum space, we perform
an inverse Mellin tranform on eq.~(\ref{four})
analytically$^2$. The final result for the resummed hadronic function,
eq.~(\ref{three}), is
\begin{equation}
W^V(\tau,Q^2)=\sum_fC_f^VA(\alpha)\int_\tau^1dz[1+{\cal H}(z,\alpha)]{d\over
dz}
\left({{\cal F}_{f{\bar f}}(\tau/z)\over z}\right),
\label{thirteen}
\end{equation}
where the parton flux is defined as
\begin{equation}
{\cal F}_{f{\bar f}}(\tau/z)=\int_0^1\delta(1-\tau/(zx_ax_b)){dx_a\over x_a}
{dx_b\over x_b}F_{f/h_1}(x_a,Q^2)F_{{\bar f}/h_2}(x_b,Q^2),
\label{fourteen}
\end{equation}
and the hard function ${\cal H}$ is given by$^{2,4}$
\begin{equation}
{\cal H}(z,\alpha)\equiv \int_0^z dz'{{\rm e}^{E\left({1\over 1-z'},\alpha
\right)}\over \pi(1-z')}\Gamma\left(1+P_1\left({1\over 1-z'},\alpha\right)
\right)\sin\left(\pi P_1\left({1\over 1-z'},\alpha\right)\right),
\label{fifteen}
\end{equation}
where $\Gamma$ is the Euler Gamma function and $P_1(x,\alpha)\equiv {\partial
\over \partial\ln x}E(x,\alpha)$.
This hard function may be used  for predicting both the
mass-distribution for the process and its derivative with respect to
rapidity or $x_F$$^5$.

\section{Comparison with experiment and conlusions}

Because of space limitation, we will treat
only a single example: the E537 fixed target
experiment at Fermilab$^7$, which observed
Drell-Yan pairs produced by antiprotons on
a tungsten target at $\sqrt{s}=15.3$ GeV
($V=\gamma$ in eqs.~(\ref{two}) and (\ref{three})).
In fig. 1, we plot the data
as well as the one-, two-loop and resummed predictions for the
mass distribution $Q^3d\sigma/ dQ$ integrated over $x_F>0$.
We used a parton distribution$^8$ based on a fit
to the {\it purely} DIS data of CDHS. Notice that
agreement between the resummation prediction and the data is acceptable,
and better than that of  the finite-order prediction.
Parton distributions with larger sea-quark content tend to
give larger predictions in the resummed cross section.

We conclude that principal-value resummation is an appropriate way to
organize
consistently large threshold corrections in pQCD to all orders
in perturbation theory,
and
that experimental results can be successfully described in this framework.
This also suggests that whatever higher-twist
part of the theory is not contained
within pQCD may be supressed, even in the
domain of phase space probed by fixed-target
experiments at momenta like those cited in fig.~1.
The inclusion of the information contained in resummation formulas
may also be important for the determination of parton distributions
by the combination of DIS and hadron-hadron data.
%\vspace{4in}
\vspace{-1.6in}
\begin{figure}[ht]
\epsfysize=375pt
\begin{center}
\hspace{-1.2in}
\leavevmode
\epsffile{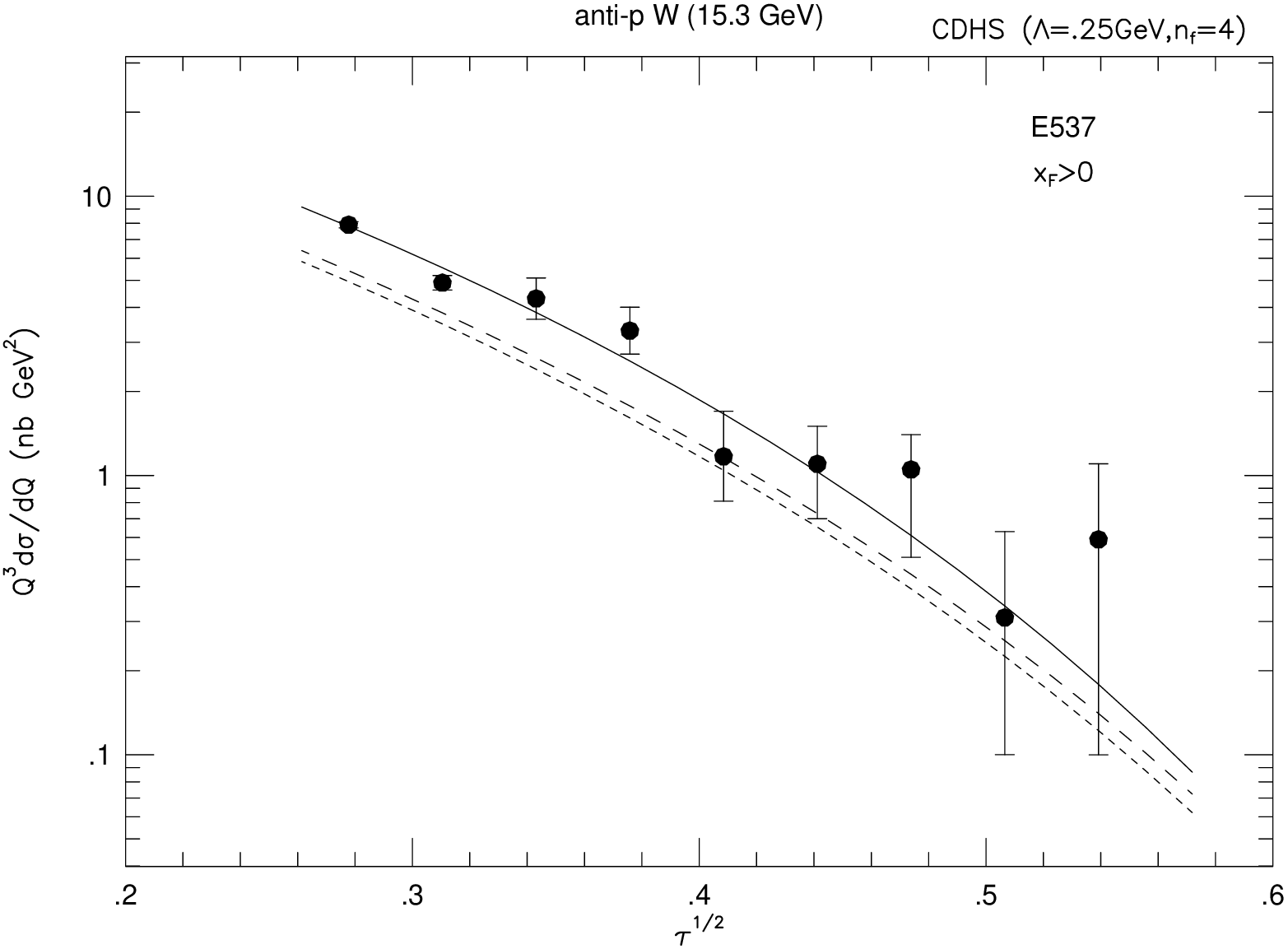}
\end{center}
\end{figure}
\vspace{-0.6in}

{\small Fig. 1. E537 experiment. Short-dash:
1-loop; Long-dash: 2-loop; Solid: Resummed.}

\section*{\normalsize{\bf Acknowledgements}}
Supported by the U.S. Department of Energy, Division of High Energy
Physics,  contract W-31-109-ENG-38,
and in part by the National Science Foundation under grant PHY 9309888 and
the Texas National Research Laboratory.

\end{document}